\documentstyle[amstex,amssymb]{article}

\catcode`\@=11
\def\numberbysection{\@addtoreset{equation}{section}
        \def\theequation{\thesection.\arabic{equation}}}
\numberbysection
\input{tcilatex}

\begin{document}

\newlength{\lno} \lno1.5cm \newlength{\len} \len=\textwidth%
\addtolength{\len}{-\lno}

\setcounter{page}{0}

\baselineskip7mm \renewcommand{\thefootnote}{\fnsymbol{footnote}} \newpage %
\setcounter{page}{0}

\begin{titlepage}     
\vspace{0.5cm}
\begin{center}
{\Large\bf Algebraic Bethe Ansatz solutions for the  $sl(2|1)^{(2)}$ and $osp(2|1)$ models with  boundary terms}\\
\vspace{1cm}
{\large  V. Kurak $^{\dag }$\hspace{.5cm} and \hspace{.5cm} A. Lima-Santos$^{\ddag}$ } \\
\vspace{1cm}
$^{\dag}${\large \em Universidade de S\~ao Paulo, Instituto de F\'{\i}sica \\
Caixa Postal 66318, CEP 05315-970~~S\~ao Paulo -SP, Brasil}\\
\vspace{.5cm}
$^{\ddag}${\large \em Universidade Federal de S\~ao Carlos, Departamento de F\'{\i}sica \\
Caixa Postal 676, CEP 13569-905~~S\~ao Carlos, Brasil}\\
\end{center}
\vspace{1.2cm}

\begin{abstract}
This work is concerned with the formulation of the graded quantum inverse scattering method  for a class of 
lattice models with reflecting boundary conditions.
The  $sl(2|1)^{(2)}$ and $osp(2|1)$ models are considered with their diagonal reflections in {\small BFB} grading.
This allowed us to derive the eigenvalues and eigenvectors for the corresponding transfer matrices as well as 
explicit expressions for the Bethe Ansatz  equations.

\end{abstract}
\vspace{2cm}
\begin{center}
\end{center}
\vfill
\begin{center}
\small{\today}
\end{center}
\end{titlepage}

\baselineskip6mm

\newpage

\section{Introduction}

Integrable quantum systems containing Fermi fields have been attracting
increasing interest due to their potential applications in condensed matter
\ physics. The prototypical examples of such systems are the supersymmetric
generalizations of the Hubbard and $t$-$J$\ models \cite{EK0}. They lead to
a generalization of the Yang-Baxter ({\small YB}) equation \cite{Baxter}
associated with the introduction of the a $Z_{2}$\ grading \cite{KS1}. In
addition to the grading, it is also important to introduce open boundary
conditions to study the boundary effects on the bulk system. The most
powerful method in the analysis of integrable models is the Bethe Ansatz (%
{\small BA}). The algebraic {\small BA, } also known as the quantum inverse
scattering method ({\small QISM}) \cite{FT}, is an elegant and important
generalization of the coordinate {\small BA} \cite{Bethe}

In this work we will study two graded three-state vertex models with
reflecting boundary conditions. Their boundary algebraic {\small BA} are
delineated based in the recent progresses \cite{GLL, KLS}, for the
formulation of the {\small QISM} for the $19$-vertex models with boundary
condition terms.

Let $V=V_{0}\oplus V_{1}$ be a $Z_{2}$-graded vector space where $0$ and $1$
denote the even and odd parts respectively. Multiplication rules in the
graded tensor product space $V\overset{s}{\otimes }V$ differ from the
ordinary ones by the appearance of additional signs. The components of a
linear operator $A\overset{s}{\otimes }B\in V\overset{s}{\otimes }V$ result
in matrix elements of the form 
\begin{equation}
(A\overset{s}{\otimes }B)_{\alpha \beta }^{\gamma \delta }=(-)^{p(\beta
)(p(\alpha )+p(\gamma ))}\ A_{\alpha \gamma }B_{\beta \delta }.
\label{int.1}
\end{equation}%
The action of the graded permutation operator ${\cal P}$ on the vector $%
\left\vert \alpha \right\rangle \overset{s}{\otimes }\left\vert \beta
\right\rangle \in V\overset{s}{\otimes }V$ is defined by 
\begin{equation}
{\cal P}\ \left\vert \alpha \right\rangle \overset{s}{\otimes }\left\vert
\beta \right\rangle =(-)^{p(\alpha )p(\beta )}\left\vert \beta \right\rangle 
\overset{s}{\otimes }\left\vert \alpha \right\rangle \Longrightarrow ({\cal P%
})_{\alpha \beta }^{\gamma \delta }=(-)^{p(\alpha )p(\beta )}\delta _{\alpha
\delta }\ \delta _{\beta \gamma }.  \label{int.2}
\end{equation}%
The graded transposition ${\rm st}$ and the graded trace {\rm str} are
defined by 
\begin{equation}
\left( A^{{\rm st}}\right) _{\alpha \beta }=(-)^{(p(\alpha )+1)p(\beta
)}A_{\beta \alpha },\quad {\rm str}A=\sum_{\alpha }(-)^{p(\alpha )}A_{\alpha
\alpha }.  \label{int.3}
\end{equation}%
where $p(\alpha )=1\ (0)$ if $\left\vert \alpha \right\rangle $ is an odd
(even) element.

For the graded case the {\small YB} equation 
\begin{equation}
{\cal R}_{12}(u){\cal R}_{13}(u+v){\cal R}_{23}(v)={\cal R}_{23}(v){\cal R}%
_{13}(u+v){\cal R}_{12}(u)  \label{int.4}
\end{equation}%
and the reflection equation \cite{Skl, MN}%
\begin{equation}
{\cal R}_{12}(u-v)K_{1}^{-}(u){\cal R}_{21}(u+v)K_{2}^{-}(v)=K_{2}^{-}(v)%
{\cal R}_{12}(u+v)K_{1}^{-}(u){\cal R}_{21}(u-v)  \label{int.5}
\end{equation}%
remain the same as in the non-graded cases and we only need to change the
usual tensor product to the graded tensor product.

In general, the dual reflection equation which depends on the unitarity and
cross-unitarity relations of the ${\cal R}$-matrix takes different forms for
different models. For the models \ considered in this paper, we write the
graded dual reflection equation in the form introduced by Zhou {\it at al}.%
\cite{Zhou} (see also \cite{FW}):%
\[
{\cal R}_{21}^{st_{1}st_{2}}(-u+v)(K_{1}^{+})^{st_{1}}(u)M_{1}^{-1}{\cal R}%
_{12}^{st_{1}st_{2}}(-u-v-2\rho )M_{1}(K_{2}^{+})^{st_{2}}(v)
\]%
\begin{equation}
=(K_{2}^{+})^{st_{2}}(v)M_{1}{\cal R}_{12}^{st_{1}st_{2}}(-u-v-2\rho
)M_{1}^{-1}(K_{1}^{+})^{st_{1}}(u){\cal R}_{21}^{st_{1}st_{2}}(-u+v),
\label{int.6}
\end{equation}%
and we will choose a common parity assignment: $p(1)=p(3)=0$ \ and $p(2)=1$
, the {\small BFB} grading.

Now, using the relations%
\begin{equation}
{\cal R}_{12}^{st_{1}st_{2}}(u)=I_{1}R_{21}(u)I_{1},\quad {\cal R}%
_{21}^{st_{1}st_{2}}(u)=I_{1}R_{12}(u)I_{1}\quad {\rm and}\quad
IK^{+}(u)I=K^{+}(u)  \label{int.7}
\end{equation}%
with $I={\rm diag}(1,-1,1)$ and the property $\left[ M_{1}M_{2},{\cal R}(u)%
\right] =0$ we can see that the usual isomorphism \cite{MN2}%
\begin{equation}
K^{-}(u):\rightarrow K^{+}(u)=K^{-}(-u-\rho )^{st}M.  \label{int.8}
\end{equation}%
holds with the {\small BFB} grading. Here ${\rm st}_{i}$ denotes
super-transposition in the space $i$.

A quantum-integrable system is characterized by the monodromy matrix $T(u)$
satisfying the fundamental relation%
\begin{equation}
R(u-v)\left[ T(u)\otimes T(v)\right] =\left[ T(v)\otimes T(u)\right] R(u-v)
\label{int.9}
\end{equation}%
where $R(u)$ is given by $R(u)=P{\cal R}(u).$

In the framework of the {\small QISM} \cite{FT}, the simplest monodromies
have become known as ${\cal L}$ operators, the Lax operators, here defined
by ${\cal L}_{aq}(u)={\cal R}_{aq}(u)$, where the subscript $a$ represents
the auxiliary space, and $q$ represents the quantum space. The monodromy
matrix $T(u)$ is defined as the matrix product of $N$ \ Lax operators on all
sites of the lattice,%
\begin{equation}
T(u)={\cal L}_{aN}(u){\cal L}_{aN-1}(u)\cdots {\cal L}_{a1}(u).
\label{int.10}
\end{equation}

The main result for open boundaries integrability is: if the boundary
equations are satisfied, then the Sklyanin's transfer matrix \cite{Skl}%
\begin{equation}
t(u)={\rm str}_{a}\left( K^{+}(u)T(u)K^{-}(u)T^{-1}(-u)\right) ,
\label{int.11}
\end{equation}%
forms a commuting collection of operators in the quantum space%
\begin{equation}
\left[ t(u),t(v)\right] =0,\qquad \forall u,v  \label{int.12}
\end{equation}

The commutativity of $t(u)$ can be proved by using the unitarity and
crossing-unitarity relations, the reflection equation and the dual
reflection equation. In particular, it implies the integrability of an open
quantum spin chain whose Hamiltonian (with $K^{-}(0)=1$) is given by \cite%
{Skl}%
\begin{equation}
H=\sum_{k=1}^{N-1}H_{k,k+1}+\frac{1}{2}\left. \frac{dK_{1}^{-}(u)}{du}%
\right\vert _{u=0}+\frac{{\rm str}_{0}K_{0}^{+}(0)H_{N,0}}{{\rm str}K^{+}(0)}%
,  \label{int.13}
\end{equation}%
where the two-site terms are given by%
\begin{equation}
H_{k,k+1}=\left. \frac{d}{du}P_{k,k+1}{\cal R}_{k,k+1}(u)\right\vert _{u=0}
\label{int.14}
\end{equation}%
in the standard fashion.

The paper is organized as follows: in Section $2$ we define the models to be
studied. In section $3$, we present our detailed calculations common to both
models and in Section $4$ the eigenspectra and the corresponding Bethe
equations are explicitly presented for each model. Section $5$ is reserved
for conclusions.

\section{The models}

The three-state vertex models that we will consider are the $sl(2|1)^{(2)}$
model and the $osp(1|2)$ model. Their ${\cal R}$-matrices have a common form%
\begin{equation}
{\cal R}(u)=\left( 
\begin{array}{ccc|ccc|ccc}
x_{1} &  &  &  &  &  &  &  &  \\ 
& x_{2} &  & x_{5} &  &  &  &  &  \\ 
&  & x_{3} &  & x_{6} &  & x_{7} &  &  \\ \hline
& y_{5} &  & x_{2} &  &  &  &  &  \\ 
&  & y_{6} &  & x_{4} &  & x_{6} &  &  \\ 
&  &  &  &  & x_{2} &  & x_{5} &  \\ \hline
&  &  &  &  &  &  &  &  \\[-10pt] 
&  & y_{7} &  & y_{6} &  & x_{3} &  &  \\ 
&  &  &  &  & y_{5} &  & x_{2} &  \\ 
&  &  &  &  &  &  &  & x_{1}%
\end{array}%
\right) ,  \label{mod.1}
\end{equation}%
satisfying the properties%
\begin{eqnarray}
{\rm regularity} &:&{\cal R}_{12}(0)=f(0)^{1/2}P_{12},  \nonumber \\
{\rm unitarity} &:&{\cal R}_{12}(u){\cal R}_{12}^{st_{1}st_{2}}(-u)=f(u), 
\nonumber \\
{\rm PT-symmetry} &:&P_{12}{\cal R}_{12}(u)P_{12}={\cal R}%
_{12}^{st_{1}st_{2}}(u),  \nonumber \\
{\rm cros}\text{sin}{\rm g-symmetry} &:&{\cal R}_{12}(u)=U_{1}{\cal R}%
_{12}^{st_{2}}(-u-\rho )U_{1}^{-1},  \label{mod.2}
\end{eqnarray}%
where $f(u)=x_{1}(u)x_{1}(-u)$, $x_{1}(u)$ being defined for each model
below. $\rho $ is the crossing parameter and $U$ determines the crossing
matrix%
\begin{equation}
M=U^{t}U=M^{t}.  \label{mod.3}
\end{equation}%
Unitarity and crossing-symmetry together imply the useful relation%
\begin{equation}
M_{1}{\cal R}_{12}^{st_{2}}(-u-\rho )M_{1}^{-1}{\cal R}_{12}^{st_{1}}(u-\rho
)=f^{^{\prime }}(u).  \label{mod.4}
\end{equation}

\subsection{{\bf The sl(2\TEXTsymbol{\vert}1)}$^{(2)}${\bf \ model}}

The solution of the graded {\small YB} equation \ corresponding to $%
sl(2|1)^{(2)}$ \ in the fundamental representation has the form (\ref{mod.1}%
) with non-zero entries \cite{KS2, BS}:%
\begin{eqnarray}
x_{1}(u) &=&\sinh (u+2\eta )\cosh (u+\eta ),\quad x_{2}(u)=\sinh u\cosh
(u+\eta ),  \nonumber \\
x_{3}(u) &=&\sinh u\cosh (u-\eta ),\quad x_{4}(u)=\sinh u\cosh (u+\eta
)-\sinh 2\eta \cosh \eta .  \nonumber \\
y_{5}(u) &=&x_{5}(u)=\sinh 2\eta \cosh (u+\eta ),\quad
y_{6}(u)=x_{6}(u)=\sinh 2\eta \sinh u,  \nonumber \\
y_{7}(u) &=&x_{7}(u)=\sinh 2\eta \cosh \eta .  \label{mod.5}
\end{eqnarray}%
This ${\cal R}$-matrix is regular and unitary, with $f(u)=x_{1}(u)x_{1}(-u)$%
, $P$- and $T$-symmetric and crossing-symmetric with $M=1$ and $\rho =\eta $%
. The graded version of the crossing-unitarity relation (\ref{mod.4}) is
satisfied with $f^{^{\prime }}(u)=x_{1}(u+i\frac{\pi }{2})x_{1}(-u-i\frac{%
\pi }{2}).$

The most general diagonal solution for $K^{-}(u)$ has been presented in Ref. 
\cite{MN3} and it is given by

\begin{equation}
K^{-}(u,\beta )=\left( 
\begin{array}{ccc}
k_{11}^{-}(u) &  &  \\ 
& 1 &  \\ 
&  & k_{33}^{-}(u)%
\end{array}%
\right) ,  \label{mod.6}
\end{equation}%
with%
\begin{equation}
k_{11}^{-}(u)=-\frac{\beta \sinh u+2\cosh u}{\beta \sinh u-2\cosh u},\quad
k_{33}^{-}(u)=\frac{\beta \cosh (u+\eta )-2\sinh (u+\eta )}{\beta \cosh
(u-\eta )+2\sinh (u-\eta )},  \label{mod.7}
\end{equation}%
where $\beta $ \ is a free parameter. Due to the automorphism (\ref{int.8})
the solution for $K^{+}(u)$ is given by $K^{-}(-u-\rho ,\frac{1}{4}\alpha )$ 
{\it i.e}.%
\begin{equation}
K^{+}(u,\alpha )=\left( 
\begin{array}{ccc}
k_{11}^{+}(u) &  &  \\ 
& 1 &  \\ 
&  & k_{33}^{+}(u)%
\end{array}%
\right) ,  \label{mod.8}
\end{equation}%
where%
\begin{equation}
k_{11}^{+}(u)=\frac{\alpha \cosh (u+\eta )-2\sinh (u+\eta )}{\alpha \cosh
(u+\eta )+2\sinh (u+\eta )},\quad k_{33}^{+}(u)=-\frac{\alpha \sinh u+2\cosh
u}{\alpha \sinh (u+2\eta )-2\cosh (u+2\eta )},  \label{mod.9}
\end{equation}%
and $\alpha $ is another free parameter.

\subsection{{\bf The osp(2\TEXTsymbol{\vert}1) model}}

The trigonometric solution of the graded {\small YB} equation \
corresponding to $osp(1|2)$ \ in the fundamental representation has the form
(\ref{mod.1}) with non-zero entries \cite{BS}:%
\begin{eqnarray}
x_{1}(u) &=&\sinh (u+2\eta )\sinh (u+3\eta ),\quad x_{2}(u)=\sinh u\sinh
(u+3\eta )  \nonumber \\
x_{3}(u) &=&\sinh u\sinh (u+\eta ),\quad x_{4}(u)=\sinh u\sinh (u+3\eta
)-\sinh 2\eta \sinh 3\eta  \nonumber \\
x_{5}(u) &=&{\rm e}^{-u}\sinh 2\eta \sinh (u+3\eta ),\quad y_{5}(u)={\rm e}%
^{u}\sinh 2\eta \sinh (u+3\eta )  \nonumber \\
x_{6}(u) &=&-{\rm e}^{-u-2\eta }\sinh 2\eta \sinh u,\quad y_{6}(u)={\rm e}%
^{u+2\eta }\sinh 2\eta \sinh u  \nonumber \\
x_{7}(u) &=&{\rm e}^{-u}\sinh 2\eta \left( \sinh (u+3\eta )+{\rm e}^{-\eta
}\sinh u\right)  \nonumber \\
y_{7}(u) &=&{\rm e}^{u}\sinh 2\eta \left( \sinh (u+3\eta )+{\rm e}^{\eta
}\sinh u\right)  \label{mod.10}
\end{eqnarray}%
This ${\cal R}$-matrix is regular and unitary, with $f^{^{\prime
}}(u)=f(u)=x_{1}(u)x_{1}(-u)$. It is $PT$-symmetric and crossing-symmetric,
with\ $\rho =3\eta $ and%
\begin{equation}
M=\left( 
\begin{array}{ccc}
{\rm e}^{-2\eta } &  &  \\ 
& 1 &  \\ 
&  & {\rm e}^{2\eta }%
\end{array}%
\right) .  \label{mod.11}
\end{equation}

Diagonal solutions for $K^{-}(u)$ have been obtained in \cite{LS1}. It turns
out that there are three solutions without free parameters, being $%
K^{-}(u)=1 $, $K^{-}(u)=F^{+}$ and $K^{-}(u)=F^{-}$, with%
\begin{equation}
F^{\pm }=\left( 
\begin{array}{ccc}
\mp {\rm e}^{-2u}f^{(\pm )}(u) &  &  \\ 
& 1 &  \\ 
&  & \mp {\rm e}^{2u}f^{(\pm )}(u)%
\end{array}%
\right) ,  \label{mod.12}
\end{equation}%
where we have defined%
\begin{equation}
f^{(+)}(u)=\frac{\sinh (u+3\eta /2)}{\sinh (u-3\eta /2)},\quad f^{(-)}(u)=%
\frac{\cosh (u+3\eta /2)}{\cosh (u-3\eta /2)}.  \label{mod.13}
\end{equation}%
By the automorphism (\ref{int.8}), three solutions $K^{+}(u)$ follow as $%
K^{+}(u)=M$, $K^{+}(u)=G^{+}$ and $K^{+}(u)=G^{-}$, with%
\begin{equation}
G^{\pm }=\left( 
\begin{array}{ccc}
\mp {\rm e}^{2u+4\eta }g^{(\pm )}(u) &  &  \\ 
& 1 &  \\ 
&  & \mp {\rm e}^{-2u-4\eta }g^{(\pm )}(u)%
\end{array}%
\right) ,  \label{mod.14}
\end{equation}%
where we have defined%
\begin{equation}
g^{(+)}(u)=\frac{\sinh (u+3\eta /2)}{\sinh (u+9\eta /2)},\quad g^{(-)}(u)=%
\frac{\cosh (u+3\eta /2)}{\cosh (u+9\eta /2)}.  \label{mod.15}
\end{equation}

We have thus nine possibilities for the commuting transfer matrix (\ref%
{int.11}). We will only consider three types of boundary solutions, one for
each pair ($K^{-}(u),K^{+}(u)$) defined by the automorphism (\ref{int.8}): $%
(1,M)$, $(F^{+},G^{+})$ and $(F^{-},G^{-})$.

\section{Algebraic Bethe Ansatz}

The monodromy matrix $T(u)$ (\ref{int.10}) \ and its reflection $T^{-1}(-u)$
\ can be written as matrices $3$ by $3$

\begin{equation}
T(u)=\left( 
\begin{array}{ccc}
T_{11}(u) & T_{12}(u) & T_{13}(u) \\ 
T_{21}(u) & T_{22}(u) & T_{23}(u) \\ 
T_{31}(u) & T_{32}(u) & T_{33}(u)%
\end{array}%
\right) ,\quad T^{-1}(-u)=\left( 
\begin{array}{ccc}
T_{11}^{-1}(-u) & T_{12}^{-1}(-u) & T_{13}^{-1}(-u) \\ 
T_{21}^{-1}(-u) & T_{22}^{-1}(-u) & T_{23}^{-1}(-u) \\ 
T_{31}^{-1}(-u) & T_{32}^{-1}(-u) & T_{33}^{-1}(-u)%
\end{array}%
\right)  \label{baba.1}
\end{equation}%
where 
\begin{equation}
T_{ia}(u)=\sum_{k_{1},...,k_{N-1}=1}^{3}{\cal L}_{ik_{1}}^{(N)}(u,\eta )%
\overset{s}{\otimes }{\cal L}_{k_{1}k_{2}}^{(N-1)}(u,\eta )\overset{s}{%
\otimes }\cdots \overset{s}{\otimes }{\cal L}_{k_{N-1}a}^{(1)}(u,\eta )
\label{baba.2}
\end{equation}%
where ${\cal L}_{ij}^{(n)}$ are $3\times 3$ matrices acting on the $n${\rm th%
} site of the lattice, defined by 
\[
{\cal L}_{11}^{(n)}=\left( 
\begin{array}{ccc}
x_{1} & 0 & 0 \\ 
0 & x_{2} & 0 \\ 
0 & 0 & x_{3}%
\end{array}%
\right) ,\quad {\cal L}_{12}^{(n)}=\left( 
\begin{array}{ccc}
0 & 0 & 0 \\ 
x_{5} & 0 & 0 \\ 
0 & x_{6} & 0%
\end{array}%
\right) ,\quad {\cal L}_{13}^{(n)}=\left( 
\begin{array}{ccc}
0 & 0 & 0 \\ 
0 & 0 & 0 \\ 
x_{7} & 0 & 0%
\end{array}%
\right) , 
\]%
\[
{\cal L}_{21}^{(n)}=\left( 
\begin{array}{ccc}
0 & y_{5} & 0 \\ 
0 & 0 & y_{6} \\ 
0 & 0 & 0%
\end{array}%
\right) ,\quad {\cal L}_{22}^{(n)}=\left( 
\begin{array}{ccc}
x_{2} & 0 & 0 \\ 
0 & x_{4} & 0 \\ 
0 & 0 & x_{2}%
\end{array}%
\right) ,\quad {\cal L}_{23}^{(n)}=\left( 
\begin{array}{ccc}
0 & 0 & 0 \\ 
x_{6} & 0 & 0 \\ 
0 & x_{5} & 0%
\end{array}%
\right) , 
\]%
\begin{equation}
{\cal L}_{31}^{(n)}=\left( 
\begin{array}{ccc}
0 & 0 & y_{7} \\ 
0 & 0 & 0 \\ 
0 & 0 & 0%
\end{array}%
\right) ,\quad {\cal L}_{32}^{(n)}=\left( 
\begin{array}{ccc}
0 & y_{6} & 0 \\ 
0 & 0 & y_{5} \\ 
0 & 0 & 0%
\end{array}%
\right) ,\quad {\cal L}_{33}^{(n)}=\left( 
\begin{array}{ccc}
x_{3} & 0 & 0 \\ 
0 & x_{2} & 0 \\ 
0 & 0 & x_{1}%
\end{array}%
\right) .  \label{baba.3}
\end{equation}%
Using the unitary relation in (\ref{mod.2}) we can see that the reflected
monodromy $T^{-1}(-u)$ has the following matrix elements%
\begin{equation}
T_{bj}^{-1}(-u)=\frac{1}{f(u)^{N}}\sum_{k_{1},...,k_{N-1}=1}^{3}{\cal L}%
_{bk_{1}}^{(1)}(-u,-\eta )\overset{s}{\otimes }{\cal L}%
_{k_{1}k_{2}}^{(2)}(-u,-\eta )\overset{s}{\otimes }\cdots \overset{s}{%
\otimes }{\cal L}_{k_{N-1}j}^{(N)}(-u,-\eta ).  \label{baba.4}
\end{equation}

For the vertex models considered in this paper we can choose the highest
weight vector of the monodromy matrix in a lattice of $N$ sites as the even
(bosonic) completely unoccupied state%
\begin{equation}
\left\vert 0\right\rangle =\dprod\limits_{k=1}^{N}\overset{s}{\otimes }%
\left\vert 0\right\rangle _{k},\qquad \left\vert 0\right\rangle _{k}=\left( 
\begin{array}{c}
1 \\ 
0 \\ 
0%
\end{array}%
\right) ,  \label{baba.5}
\end{equation}%
where $\left\vert 0\right\rangle _{k}$ is the local reference state at the $%
k $-{\it th} lattice site with three components

The action of $T(u)$ and $T^{-1}(-u)$ on this state are 
\begin{equation}
T(u)\left\vert 0\right\rangle =f^{N}(u)T^{-1}(-u)\left\vert 0\right\rangle
=\left( 
\begin{array}{ccc}
x_{1}^{N}(u)\left\vert 0\right\rangle & \ast & \ast \ast \\ 
0 & x_{2}^{N}(u)\left\vert 0\right\rangle & \ast \ast \ast \\ 
0 & 0 & x_{3}^{N}(u)\left\vert 0\right\rangle%
\end{array}%
\right) ,  \label{baba.6}
\end{equation}%
which give us, in the usual {\small BA} language, creation and annihilation
operators for (\ref{baba.5}). Moreover, we are working out with boundaries
and in this case we have a double-row monodromy defined by%
\begin{equation}
U(u)=T(u)K^{-}(u)T^{-1}(-u)=\left( 
\begin{array}{ccc}
U_{11}(u) & U_{12}(u) & U_{13}(u) \\ 
U_{21}(u) & U_{22}(u) & U_{23}(u) \\ 
U_{31}(u) & U_{32}(u) & U_{33}(u)%
\end{array}%
\right)  \label{baba.7}
\end{equation}%
where $K^{(-)}(u)$ is a reflection matrix.

For $K^{(-)}(u)={\rm diag}(k_{11}^{-}(u),k_{22}^{-}(u),k_{33}^{-}(u))$, the
matrix elements of $U$ have the form 
\begin{equation}
U_{ij}(u)=\sum_{a=1}^{3}T_{ia}(u)k_{aa}^{-}(u)T_{aj}^{-1}(-u),\qquad
i,j=1,2,3.  \label{baba.8}
\end{equation}%
It follows from (\ref{baba.8}) that we will need to know the commutation
relations between the operators $T(u)$ and $T^{-1}(-u)$ in order to get the
action of $U(u)$ on the reference state (\ref{baba.5}). Using the
fundamental relation (\ref{int.9}) with $u=-v$ we will get the matrix
relation%
\begin{equation}
T_{2}^{-1}(-u)R_{12}(2u)T_{1}(u)=T_{1}(u)R_{12}(2u)T_{2}^{-1}(-u)
\label{baba.9}
\end{equation}%
Applying both sides of this relation on the reference state, we find the
following relations%
\begin{equation}
T_{21}(u)T_{12}^{-1}(-u)\left\vert 0\right\rangle =f_{1}(u)\frac{%
x_{1}^{2N}(u)-x_{2}^{2N}(u)}{f^{N}(u)}\left\vert 0\right\rangle
\label{baba.10}
\end{equation}%
\begin{equation}
T_{31}(u)T_{13}^{-1}(-u)\left\vert 0\right\rangle =\left\{ f_{2}(u)\frac{%
x_{1}^{2N}(u)}{f^{N}(u)}-f_{3}(u)f_{1}(u)\frac{x_{2}^{2N}(u)}{f^{N}(u)}%
-f_{4}(u)\frac{x_{3}^{2N}(u)}{f^{N}(u)}\right\} \left\vert 0\right\rangle
\label{baba.11}
\end{equation}%
\begin{equation}
T_{32}(u)T_{23}^{-1}(-u)\left\vert 0\right\rangle =f_{2}(u)\frac{%
x_{2}^{2N}(u)-x_{3}^{2N}(u)}{f^{N}(u)}\left\vert 0\right\rangle
\label{baba.12}
\end{equation}%
where%
\begin{eqnarray}
f_{1}(u) &=&\frac{y_{5}(2u)}{x_{1}(2u)},\qquad f_{2}(u)=\frac{y_{7}(2u)}{%
x_{1}(2u)},\qquad  \nonumber \\
f_{3}(u) &=&-\frac{x_{1}(2u)y_{5}(2u)-x_{5}(2u)y_{7}(2u)}{%
x_{1}(2u)x_{4}(2u)+x_{5}(2u)y_{5}(2u)},  \nonumber \\
f_{4}(u) &=&\frac{x_{4}(2u)y_{7}(2u)+y_{5}^{2}(2u)}{%
x_{1}(2u)x_{4}(2u)+x_{5}(2u)y_{5}(2u)}.  \label{baba.13}
\end{eqnarray}%
Using these relations we can get the action of each operator $U_{ij}$ on the
reference state: for the diagonal entries we have%
\begin{eqnarray}
U_{11}(u)\left\vert 0\right\rangle &=&k_{11}^{-}(u)\frac{x_{1}^{2N}(u)}{%
f^{N}(u)}\left\vert 0\right\rangle  \nonumber \\
&&  \nonumber \\
U_{22}(u)\left\vert 0\right\rangle &=&f_{1}(u)U_{11}(u)\left\vert
0\right\rangle +\left[ k_{22}^{-}(u)-k_{11}^{-}(u)f_{1}(u)\right] \frac{%
x_{2}^{2N}(u)}{f^{N}(u)}\left\vert 0\right\rangle  \nonumber \\
&&  \nonumber \\
U_{33}(u)\left\vert 0\right\rangle &=&\left[ \left(
f_{2}(u)-f_{1}(u)f_{3}(u)\right) U_{11}(u)+f_{3}(u)U_{22}(u)\right]
\left\vert 0\right\rangle  \nonumber \\
&&+\left[ k_{33}^{-}(u)-k_{22}^{-}(u)f_{3}(u)-k_{11}^{-}(u)f_{4}(u)\right] 
\frac{x_{3}^{2N}(u)}{f^{N}(u)}\left\vert 0\right\rangle  \label{baba.14}
\end{eqnarray}%
and for the elements out of the diagonal we get 
\begin{equation}
U_{ij}(u)\left\vert 0\right\rangle =0,\quad (i>j),\qquad U_{ij}(u)\left\vert
0\right\rangle \neq \left\{ 0,\left\vert 0\right\rangle \right\} ,\quad (i<j)
\label{baba.15}
\end{equation}

Now, we define news operators:%
\[
{\cal D}_{1}(u)=U_{11}(u),\qquad {\cal B}_{1}(u)=U_{12}(u),\qquad {\cal B}%
_{2}(u)=U_{13}(u) 
\]%
\[
{\cal C}_{1}(u)=U_{21}(u),\qquad {\cal D}_{2}(u)=U_{22}(u)-f_{1}(u){\cal D}%
_{1}(u),\qquad {\cal B}_{3}(u)=U_{23}(u) 
\]%
\begin{equation}
{\cal C}_{2}(u)=U_{31}(u),\qquad {\cal C}_{3}(u)=U_{32}(u),\qquad {\cal D}%
_{3}(u)=U_{33}(u)-f_{2}(u){\cal D}_{1}(u)-f_{3}(u){\cal D}_{2}(u)
\label{baba.16}
\end{equation}%
to write the double-row monodromy matrix in the form%
\begin{equation}
U(u)\rightarrow {\cal U}(u)=\left( 
\begin{array}{ccc}
{\cal D}_{1}(u) & {\cal B}_{1}(u) & {\cal B}_{2}(u) \\ 
{\cal C}_{1}(u) & {\cal D}_{2}(u) & {\cal B}_{3}(u) \\ 
{\cal C}_{2}(u) & {\cal C}_{3}(u) & {\cal D}_{3}(u)%
\end{array}%
\right) .  \label{baba.17}
\end{equation}

The action of $\ {\cal U}(u)$ on the reference state has the usual {\small BA%
} form 
\begin{equation}
{\cal U}(u)\left\vert 0\right\rangle =\left( 
\begin{array}{ccc}
{\cal X}_{1}(u)\left\vert 0\right\rangle & \ast & \ast \ast \\ 
0 & {\cal X}_{2}(u)\left\vert 0\right\rangle & \ast \ast \ast \\ 
0 & 0 & {\cal X}_{3}(u)\left\vert 0\right\rangle%
\end{array}%
\right)  \label{baba.18}
\end{equation}%
where%
\begin{eqnarray}
{\cal X}_{1}(u) &=&k_{11}^{-}(u)\frac{x_{1}^{2N}(u)}{f^{N}(u)}  \nonumber \\
{\cal X}_{2}(u) &=&\left[ k_{22}^{-}(u)-k_{11}^{-}(u)f_{1}(u)\right] \frac{%
x_{2}^{2N}(u)}{f^{N}(u)}  \nonumber \\
{\cal X}_{3}(u) &=&\left[
k_{33}^{-}(u)-k_{22}^{-}(u)f_{3}(u)-k_{11}^{-}(u)f_{4}(u)\right] \frac{%
x_{3}^{2N}(u)}{f^{N}(u)}  \label{baba.19}
\end{eqnarray}

The transfer matrix \ $t(u)={\rm str}(K^{+}U)$, with diagonal left
reflection \ $K^{(+)}={\rm diag}(k_{11}^{+},k_{22}^{+}.k_{33}^{+})$ and 
{\small BFB} grading, has the form%
\begin{eqnarray}
t(u) &=&k_{11}^{+}(u)U_{11}(u)-k_{22}^{+}(u)U_{22}(u)+k_{33}^{+}(u)U_{33}(u)
\nonumber \\
&=&\Omega _{1}(u){\cal D}_{1}(u)+\Omega _{2}(u){\cal D}_{2}(u)+\Omega _{3}(u)%
{\cal D}_{3}(u)  \label{baba.20}
\end{eqnarray}%
where%
\begin{eqnarray}
\Omega _{1}(u) &=&k_{11}^{+}(u)-f_{1}(u)k_{22}^{+}(u)+f_{2}(u)k_{33}^{+}(u) 
\nonumber \\
\Omega _{2}(u) &=&-k_{22}^{+}(u)+f_{3}(u)k_{33}^{+}(u)  \nonumber \\
\Omega _{3}(u) &=&k_{33}^{+}(u)  \label{baba.21}
\end{eqnarray}%
Here we note the sign ($-$) from the graded trace is absorbed in the
definition of $\Omega _{2}(u)$.

From ${\cal U}(u)$ follows the usual algebraic {\small BA} structure.
Therefore we can look for states created by the operators ${\cal B}_{i}(u)$
from a reference $\Psi _{0}$ which will be eigenstates of (\ref{baba.20}).
To do this we first recall the magnon number operator%
\begin{equation}
M=\sum_{k=1}^{N}M_{k},\qquad M_{k}={\rm diag}(0,1,2)  \label{baba.22}
\end{equation}%
This is the analogue of the operator $S_{T}^{z}$ used in the coordinate 
{\small BA} construction. The relation $m=N-S_{T}^{z}$ \ ($M\Psi _{m}=m\Psi
_{m}$ ) allows us build states $\Psi _{m}$ such that $t(u)\Psi _{m}=\Lambda
_{m}\Psi _{m}$. Therefore, we can start the diagonalization of $t(u)$ by
considering all possible values of $m$ in a lattice with $N$ sites.

By the previous construction, $\Psi _{0}$ \ is our reference state $|0>$ ,
which is itself an eigenstate of $t(u)$ 
\begin{equation}
t(u)\Psi _{0}=\Lambda _{0}(u)\Psi _{0}  \label{baba.30}
\end{equation}%
with eigenvalue%
\begin{eqnarray}
\Lambda _{0}(u) &=&\left[
k_{11}^{+}(u)-f_{1}(u)k_{22}^{+}(u)+f_{2}(u)k_{33}^{+}(u)\right]
k_{11}^{-}(u)\frac{x_{1}^{2N}(u)}{f^{N}(u)}  \nonumber \\
&&+\left[ -k_{22}^{+}(u)+f_{3}(u)k_{33}^{+}(u)\right] \left[
k_{22}^{-}(u)-k_{11}^{-}(u)f_{1}(u)\right] \frac{x_{2}^{2N}(u)}{f^{N}(u)} 
\nonumber \\
&&+k_{33}^{+}(u)\left[
k_{33}^{-}(u)-k_{22}^{-}(u)f_{3}(u)-k_{11}^{-}(u)f_{4}(u)\right] \frac{%
x_{3}^{2N}(u)}{f^{N}(u)}  \label{baba.31}
\end{eqnarray}%
It is the only state with $m=0$.

\subsection{The one-particle state}

For $m=1$ we seek a state of the form%
\begin{equation}
\Psi _{1}(u_{1})={\cal B}_{1}(u_{1})\left\vert 0\right\rangle .
\label{baba.32}
\end{equation}

The action of $t(u)$ on this state is given by%
\begin{equation}
t(u)\Psi _{1}(u_{1})=\Omega _{1}(u){\cal D}_{1}(u){\cal B}%
_{1}(u_{1})\left\vert 0\right\rangle +\Omega _{2}(u){\cal D}_{2}(u){\cal B}%
_{1}(u_{1})\left\vert 0\right\rangle +\Omega _{3}(u){\cal D}_{3}(u){\cal B}%
_{1}(u_{1})\left\vert 0\right\rangle .  \label{baba.33}
\end{equation}%
Since we know from (\ref{baba.18}) the action of the operators ${\cal D}%
_{i}(u)$ on the reference state $\left\vert 0\right\rangle $, we need to
arrange the operators products 
\begin{equation}
{\cal D}_{1}(u){\cal B}_{1}(u_{1}),\ \quad {\cal D}_{2}(u){\cal B}%
_{1}(u_{1})\quad {\rm and}\quad {\cal D}_{3}(u){\cal B}_{1}(u_{1})
\label{baba.34}
\end{equation}%
in a normal-ordered form \cite{TA}: We anticipate that, in general, the
operator-valued function $\Psi _{n}(u_{1},\ldots ,u_{n})$ for a n-particle
Bethe state will be composed by a set of normal-ordered monomials. A
monomial is said to be in normal order if all elements ${\cal B}_{i}$ are on
the left, and all elements ${\cal C}_{i}$ are on the right of the elements $%
{\cal D}_{i}.$

In order to get this normal ordering we recall that the double-row monodromy
matrix ${\cal U}(u)$ satisfies the fundamental reflection equation 
\begin{equation}
{\cal R}_{12}(u-v){\cal U}_{1}(u){\cal R}_{21}(u+v){\cal U}_{2}(v)={\cal U}%
_{2}(v){\cal R}_{12}(u+v){\cal U}_{1}(u){\cal R}_{21}(u-v),  \label{baba.35}
\end{equation}%
where ${\cal U}_{1}(u)={\cal U}(u)\overset{s}{\otimes }1,\ {\cal U}_{2}(u)=1%
\overset{s}{\otimes }{\cal U}(u)$ and ${\cal R}_{21}(u)={\cal PR}_{12}(u)%
{\cal P}$.

In the appendix of \cite{KLS} it was shown as this equation can be used to
recast the non-normal ordered products as the above into a linear
combination of normal-ordered ones. However, in this article we will present
only the main results. (See below)

For the present case $t(u)\Psi _{1}(u_{1})$ can be computed with the aid of
the following normal-ordered (or commutation) relations%
\begin{eqnarray}
{\cal D}_{1}(u){\cal B}_{1}(u_{1}) &=&a_{11}(u,u_{1}){\cal B}_{1}(u_{1})%
{\cal D}_{1}(u)+a_{12}(u,u_{1}){\cal B}_{1}(u){\cal D}%
_{1}(u_{1})+a_{13}(u,u_{1}){\cal B}_{1}(u){\cal D}_{2}(u_{1})  \nonumber \\
&&+a_{14}(u,u_{1}){\cal B}_{2}(u){\cal C}_{1}(u_{1})+a_{15}(u,u_{1}){\cal B}%
_{2}(u){\cal C}_{3}(u_{1})+a_{16}(u,u_{1}){\cal B}_{2}(u_{1}){\cal C}_{1}(u)
\nonumber \\
&&
\end{eqnarray}%
\begin{eqnarray}
{\cal D}_{2}(u){\cal B}_{1}(u_{1}) &=&a_{21}(u,u_{1}){\cal B}_{1}(u_{1})%
{\cal D}_{2}(u)+a_{22}(u,u_{1}){\cal B}_{1}(u){\cal D}%
_{1}(u_{1})+a_{23}(u,u_{1}){\cal B}_{1}(u){\em D}_{2}(u_{1})  \nonumber \\
&&+a_{24}(u,u_{1}){\cal B}_{3}(u){\cal D}_{1}(u_{1})+a_{25}(u,u_{1}){\cal B}%
_{3}(u){\cal D}_{2}(u_{1})+a_{26}(u,u_{1}){\cal B}_{2}(u){\cal C}_{1}(u_{1})
\nonumber \\
&&+a_{27}(u,u_{1}){\cal B}_{2}(u){\cal C}_{3}(u_{1})+a_{28}(u,u_{1}){\cal B}%
_{2}(u_{1}){\cal C}_{1}(u)+a_{29}(u,u_{1}){\cal B}_{2}(u_{1}){\cal C}_{3}(u)
\nonumber \\
&&  \label{baba.37}
\end{eqnarray}%
\begin{eqnarray}
{\cal D}_{3}(u){\cal B}_{1}(u_{1}) &=&a_{31}(u,u_{1}){\cal B}_{1}(u_{1})%
{\cal D}_{3}(u)+a_{32}(u,u_{1}){\cal B}_{1}(u){\cal D}%
_{1}(u_{1})+a_{33}(u,u_{1}){\cal B}_{1}(u){\cal D}_{2}(u_{1})  \nonumber \\
&&+a_{34}(u,u_{1}){\cal B}_{3}(u){\cal D}_{1}(u_{1})+a_{35}(u,u_{1}){\cal B}%
_{3}(u){\cal D}_{2}(u_{1})+a_{36}(u,u_{1}){\cal B}_{2}(u){\cal C}_{1}(u_{1})
\nonumber \\
&&+a_{37}(u,u_{1}){\cal B}_{2}(u){\cal C}_{3}(u_{1})+a_{38}(u,u_{1}){\cal B}%
_{2}(u_{1}){\cal C}_{1}(u)+a_{39}(u,u_{1}){\cal B}_{2}(u_{1}){\cal C}_{3}(u)
\nonumber \\
&&  \label{baba.38}
\end{eqnarray}%
Substituting these relations into (\ref{baba.33}) one gets%
\begin{eqnarray}
t(u)\Psi _{1}(u_{1}) &=&\Omega _{1}(u){\cal D}_{1}(u){\cal B}%
_{1}(u_{1})\left\vert 0\right\rangle +\Omega _{2}(u){\cal D}_{2}(u){\cal B}%
_{1}(u_{1})\left\vert 0\right\rangle +\Omega _{3}(u){\cal D}_{3}(u){\cal B}%
_{1}(u_{1})\left\vert 0\right\rangle  \nonumber \\
&=&\left[ a_{11}(u,u_{1})\Omega _{1}(u){\cal X}_{1}(u)+a_{21}(u,u_{1})\Omega
_{2}(u){\cal X}_{2}(u)+a_{31}(u,u_{1})\Omega _{3}(u){\cal X}_{3}(u)\right]
\Psi _{1}(u_{1})  \nonumber \\
&&+[{\cal X}_{1}(u_{1})\sum_{j=1}^{3}\Omega _{j}(u)a_{j2}(u,u_{1})+{\cal X}%
_{2}(u_{1})\sum_{j=1}^{3}\Omega _{j}(u)a_{j3}(u,u_{1})]B_{1}(u)\left\vert
0\right\rangle  \nonumber \\
&&+[{\cal X}_{1}(u_{1})\sum_{j=2}^{3}\Omega _{j}(u)a_{j4}(u,u_{1})+{\cal X}%
_{2}(u_{1})\sum_{j=2}^{3}\Omega _{j}(u)a_{j5}(u,u_{1})]B_{3}(u)\left\vert
0\right\rangle  \label{baba.39}
\end{eqnarray}%
So, $\Psi _{1}(u_{1})$ will be an eigenstate of $t(u)$ with eigenvalue%
\begin{equation}
\Lambda _{1}(u,u_{1})=\sum_{j=1}^{3}\Omega _{j}(u){\cal X}%
_{j}(u)a_{j1}(u,u_{1})  \label{baba.40}
\end{equation}%
provided the following equation is satisfied%
\begin{equation}
\frac{{\cal X}_{1}(u_{1})}{{\cal X}_{2}(u_{1})}=-\frac{\sum_{j=1}^{3}\Omega
_{j}(u)a_{j3}(u,u_{1})}{\sum_{j=1}^{3}\Omega _{j}(u)a_{j2}(u,u_{1})}=-\frac{%
\sum_{j=2}^{3}\Omega _{j}(u)a_{j5}(u,u_{1})}{\sum_{j=2}^{3}\Omega
_{j}(u)a_{j4}(u,u_{1})}\equiv \Theta (u_{1}).  \label{baba.41}
\end{equation}%
Explicit calculations of these expressions for the models considered in this
paper will be presented in the next section.

\subsection{The two-particle state}

For $m=2$ we have to seek eigenstates of $t(u)$ in the form%
\begin{equation}
\Psi _{2}(u_{1},u_{2})={\cal B}_{1}(u_{1}){\cal B}_{1}(u_{2})\left\vert
0\right\rangle +{\cal B}_{2}(u_{1})\Gamma (u_{1},u_{2})\left\vert
0\right\rangle  \label{baba.42}
\end{equation}%
where $\Gamma (u_{1},u_{2})$ is an operator-valued function. Next we will
use the condition that $\Psi _{2}(u_{1},u_{2})$ must be normal-ordered to
find $\Gamma (u_{1},u_{2})$.

The first term of the right hand side of (\ref{baba.42}) has its
normal-ordered form gives by the commutations relations:%
\begin{eqnarray}
{\cal B}_{1}(u_{1}){\cal B}_{1}(u_{2}) &=&\omega (u_{1},u_{2})\left[ {\cal B}%
_{1}(u_{2}){\cal B}_{1}(u_{1})+G_{d_{1}}(u_{2},u_{1}){\cal B}_{2}(u_{2})%
{\cal D}_{1}(u_{1})+G_{d_{2}}(u_{2},u_{1}){\cal B}_{2}(u_{2}){\cal D}%
_{2}(u_{1})\right]  \nonumber \\
&&-G_{d_{1}}(u_{1},u_{2}){\cal B}_{2}(u_{1}){\cal D}%
_{1}(u_{2})-G_{d_{2}}(u_{1},u_{2}){\cal B}_{2}(u_{1}){\cal D}_{2}(u_{2})
\label{baba.43}
\end{eqnarray}%
where%
\begin{equation}
\omega (u_{1},u_{2})=-\frac{%
x_{3}(u_{1}-u_{2})x_{4}(u_{1}-u_{2})-x_{6}(u_{1}-u_{2})y_{6}(u_{1}-u_{2})}{%
x_{1}(u_{1}-u_{2})x_{3}(u_{1}-u_{2})}  \label{baba.44}
\end{equation}%
\begin{equation}
G_{d_{1}}(u_{1},u_{2})=\frac{x_{6}(u_{1}-u_{2})}{x_{3}(u_{1}-u_{2})}\frac{%
x_{2}(2u_{2})}{x_{1}(2u_{2})}  \label{baba.45}
\end{equation}%
\begin{equation}
G_{d_{2}}(u_{1},u_{2})=-\frac{x_{6}(u_{1}+u_{2})}{x_{2}(u_{1}+u_{2})}
\label{baba.46}
\end{equation}%
Here we have used the following identities valid for both models,%
\begin{equation}
\frac{y_{6}(-u)}{x_{3}(-u)}=\frac{x_{3}(u)x_{6}(u)-x_{7}(u)y_{6}(u)}{%
x_{3}(u)x_{4}(u)-x_{6}(u)y_{6}(u)}  \label{baba.47}
\end{equation}%
\begin{equation}
\frac{x_{2}(2u)}{x_{1}(2u)}=\frac{%
y_{5}(u-v)x_{2}(u+v)+x_{2}(u-v)x_{5}(u+v)f_{1}(u)}{y_{5}(u-v)x_{1}(u+v)}
\label{baba.48}
\end{equation}%
Now we can see that (\ref{baba.42}) is normally ordered if it satisfies the
condition%
\begin{equation}
\Psi _{2}(u_{2},u_{1})=\omega (u_{2},u_{1})\Psi _{2}(u_{1},u_{2})
\label{baba.49}
\end{equation}%
This condition fixes $\Gamma (u_{1},u_{2})$ and, by construction, the unique
candidate for eigenstate of $t(u)$ in the case $m=2$ \ has the form%
\begin{equation}
\Psi _{2}(u_{1},u_{2})={\cal B}_{1}(u_{1}){\cal B}_{1}(u_{2})\left\vert
0\right\rangle +G_{d_{1}}(u_{1},u_{2}){\cal B}_{2}(u_{1}){\cal D}%
_{1}(u_{2})\left\vert 0\right\rangle +G_{d_{2}}(u_{1},u_{2}){\cal B}%
_{2}(u_{1}){\cal D}_{2}(u_{2})\left\vert 0\right\rangle  \label{baba.50}
\end{equation}

The action of $t(u)$ on this state in its normal ordered form is obtained
with the following commutation relations, in addition to those presented in
the $m=1$ case:%
\begin{eqnarray}
{\cal D}_{1}(u){\cal B}_{2}(v) &=&b_{11}(u,v){\cal B}_{2}(v){\cal D}%
_{1}(u)+b_{12}(u,v){\cal B}_{2}(u){\cal D}_{1}(v)+b_{13}(u,v){\cal B}_{2}(u)%
{\cal D}_{2}(v)  \nonumber \\
&&+b_{14}(u,v){\cal B}_{2}(u){\cal D}_{3}(v)+b_{15}(u,v){\cal B}_{1}(u){\cal %
B}_{1}(v)+b_{16}(u,v){\cal B}_{1}(u){\cal B}_{3}(v)  \label{baba.51}
\end{eqnarray}%
\begin{eqnarray}
{\cal D}_{2}(u){\cal B}_{2}(v) &=&b_{21}(u,v){\cal B}_{2}(v){\cal D}%
_{2}(u)+b_{22}(u,v){\cal B}_{2}(u){\cal D}_{1}(v)+b_{23}(u,v){\cal B}_{2}(u)%
{\cal D}_{2}(v)  \nonumber \\
&&+b_{24}(u,v){\cal B}_{2}(u){\cal D}_{3}(v)+b_{25}(u,v){\cal B}_{1}(u){\cal %
B}_{1}(v)+b_{26}(u,v){\cal B}_{1}(u){\cal B}_{3}(v)  \nonumber \\
&&+b_{27}(u,v){\cal B}_{3}(u){\cal B}_{1}(v)+b_{28}(u,v){\cal B}_{3}(u){\cal %
B}_{3}(v)  \label{baba.52}
\end{eqnarray}%
\begin{eqnarray}
{\cal D}_{3}(u){\cal B}_{2}(v) &=&b_{31}(u,v){\cal B}_{2}(v){\cal D}%
_{3}(u)+b_{32}(u,v){\cal B}_{2}(u){\cal D}_{1}(v)+b_{33}(u,v){\cal B}_{2}(u)%
{\cal D}_{2}(v)  \nonumber \\
&&+b_{34}(u,v){\cal B}_{2}(u){\cal D}_{3}(v)+b_{35}(u,v){\cal B}_{1}(u){\cal %
B}_{1}(v)+b_{36}(u,v){\cal B}_{1}(u){\cal B}_{3}(v)  \nonumber \\
&&+b_{37}(u,v){\cal B}_{3}(u){\cal B}_{1}(v)+b_{38}(u,v){\cal B}_{3}(u){\cal %
B}_{3}(v)  \label{baba.53}
\end{eqnarray}%
\begin{eqnarray}
{\cal C}_{1}(u){\cal B}_{1}(v) &=&c_{11}(u,v){\cal B}_{1}(v){\cal C}%
_{1}(u)+c_{12}(u,v){\cal B}_{1}(v){\cal C}_{3}(v)+c_{13}(u,v){\cal B}_{1}(u)%
{\cal C}_{3}(v)  \nonumber \\
&&+c_{14}(u,v){\cal B}_{2}(u){\cal C}_{3}(v)+c_{15}(u,v){\cal B}_{2}(v){\cal %
C}_{2}(u)+c_{16}(u,v){\cal D}_{1}(v){\cal D}_{1}(u)  \nonumber \\
&&+c_{17}(u,v){\cal D}_{1}(v){\cal D}_{2}(u)+c_{18}(u,v){\cal D}_{1}(u){\cal %
D}_{1}(v)+c_{19}(u,v){\cal D}_{1}(u){\cal D}_{2}(v)  \nonumber \\
&&+c_{110}(u,v){\cal D}_{2}(u){\cal D}_{1}(v)+c_{111}(u,v){\cal D}_{2}(u)%
{\cal D}_{2}(v)  \label{baba.54}
\end{eqnarray}%
\begin{eqnarray}
{\cal C}_{3}(u){\cal B}_{1}(v) &=&c_{21}(u,v){\cal B}_{1}(v){\cal C}%
_{1}(u)+c_{22}(u,v){\cal B}_{1}(v){\cal C}_{3}(v)+c_{23}(u,v){\cal B}_{1}(u)%
{\cal C}_{3}(v)  \nonumber \\
&&+c_{24}(u,v){\cal B}_{2}(u){\cal C}_{3}(v)+c_{25}(u,v){\cal B}_{2}(v){\cal %
C}_{2}(u)+c_{26}(u,v){\cal D}_{1}(v){\cal D}_{1}(u)  \nonumber \\
&&+c_{27}(u,v){\cal D}_{1}(v){\cal D}_{2}(u)+c_{28}(u,v){\cal D}_{1}(u){\cal %
D}_{3}(v)+c_{29}(u,v){\cal D}_{1}(u){\cal D}_{2}(v)  \nonumber \\
&&+c_{210}(u,v){\cal D}_{1}(u){\cal D}_{2}(v)+c_{211}(u,v){\cal D}_{2}(u)%
{\cal D}_{1}(v)+c_{212}(u,v){\cal D}_{2}(u){\cal D}_{2}(v)  \nonumber \\
&&+c_{213}(u,v){\cal D}_{3}(u){\cal D}_{1}(v)+c_{214}(u,v){\cal D}_{3}(u)%
{\cal D}_{1}(v))  \label{baba.55}
\end{eqnarray}%
After a straightforward calculation we obtain%
\[
t(u)\Psi _{2}(u_{1},u_{2})=[\sum_{j=1}^{3}\Omega _{j}(u){\cal X}%
_{j}(u)a_{j1}(u,u_{1})a_{j1}(u,u_{2})]\Psi _{2}(u_{1},u_{2}) 
\]%
\[
+[a_{11}(u_{1},u_{2}){\cal X}_{1}(u_{1})\sum_{j=1}^{3}\Omega
_{j}(u)a_{j2}(u,u_{1})+a_{21}(u_{1},u_{2}){\cal X}_{2}(u_{1})\sum_{j=1}^{3}%
\Omega _{j}(u)a_{j3}(u,u_{1})]{\cal B}_{1}(u){\cal B}_{1}(u_{2})\left\vert
0\right\rangle 
\]%
\[
+[a_{11}(u_{1},u_{2}){\cal X}_{1}(u_{1})\sum_{j=2}^{3}\Omega
_{j}(u)a_{j4}(u,u_{1})+a_{21}(u_{1},u_{2}){\cal X}_{2}(u_{1})\sum_{j=2}^{3}%
\Omega _{j}(u)a_{j5}(u,u_{1})]{\cal B}_{3}(u){\cal B}_{1}(u_{2})\left\vert
0\right\rangle 
\]%
\[
+[a_{11}(u_{2},u_{1}){\cal X}_{1}(u_{2})\sum_{j=1}^{3}\Omega
_{j}(u)a_{j2}(u,u_{2})+a_{21}(u_{2},u_{1}){\cal X}_{2}(u_{2})\sum_{j=1}^{3}%
\Omega _{j}(u)a_{j3}(u,u_{2})]\omega (u_{1},u_{2}){\cal B}_{1}(u){\em B}%
_{1}(u_{1})\left\vert 0\right\rangle 
\]%
\[
+[a_{11}(u_{2},u_{1}){\cal X}_{1}(u_{2})\sum_{j=2}^{3}\Omega
_{j}(u)a_{j4}(u,u_{2})+a_{21}(u_{2},u_{1}){\cal X}_{2}(u_{2})\sum_{j=2}^{3}%
\Omega _{j}(u)a_{j5}(u,u_{2})]\omega (u_{1},u_{2}){\cal B}_{3}(u){\cal B}%
_{1}(u_{1})\left\vert 0\right\rangle 
\]%
\begin{eqnarray}
&&+[{\cal X}_{1}(u_{1}){\cal X}_{1}(u_{2})\sum_{j=1}^{3}\Omega
_{j}(u)H_{j1}(u_{1},u_{2})+{\cal X}_{1}(u_{1}){\cal X}_{2}(u_{2})%
\sum_{j=1}^{3}\Omega _{j}(u)H_{j3}(u_{1},u_{2})  \nonumber \\
&&+{\cal X}_{2}(u_{1}){\cal X}_{1}(u_{2})\sum_{j=1}^{3}\Omega
_{j}(u)H_{j2}(u_{1},u_{2})+{\cal X}_{2}(u_{1}){\cal X}_{2}(u_{2})%
\sum_{j=1}^{3}\Omega _{j}(u)H_{j4}(u_{1},u_{2})]{\cal B}_{2}(u)\left\vert
0\right\rangle  \label{baba.56}
\end{eqnarray}%
where%
\begin{eqnarray}
H_{11}(u_{1},u_{2}) &=&a_{14}(u,u_{1})\left(
c_{16}(u_{1},u_{2})+c_{18}(u_{1},u_{2})\right) +a_{15}(u,u_{1})\left(
c_{26}(u_{1},u_{2})+c_{29}(u_{1},u_{2})\right)  \nonumber \\
&&+b_{12}(u,u_{1})G_{d_{1}}(u_{1},u_{2})+\omega
(u_{1},u)a_{11}(u,u_{1})a_{12}(u,u_{2})G_{d_{1}}(u,u_{1})  \nonumber \\
H_{12}(u_{1},u_{2}) &=&a_{14}(u,u_{1})\left(
c_{17}(u_{1},u_{2})+c_{110}(u_{1},u_{2})\right) +a_{15}(u,u_{1})\left(
c_{27}(u_{1},u_{2})+c_{211}(u_{1},u_{2})\right)  \nonumber \\
&&+b_{13}(u,u_{1})G_{d_{1}}(u_{1},u_{2})+\omega
(u_{1},u)a_{11}(u,u_{1})a_{12}(u,u_{2})G_{d_{2}}(u,u_{1})  \nonumber \\
H_{13}(u_{1},u_{2})
&=&a_{14}(u,u_{1})c_{19}(u_{1},u_{2})+a_{15}(u,u_{1})c_{210}(u_{1},u_{2})+b_{12}(u,u_{1})G_{d_{2}}(u_{1},u_{2})
\nonumber \\
&&+\omega (u_{1},u)a_{11}(u,u_{1})a_{13}(u,u_{2})G_{d_{1}}(u,u_{1}) 
\nonumber \\
H_{14}(u_{1},u_{2})
&=&a_{14}(u,u_{1})c_{111}(u_{1},u_{2})+a_{15}(u,u_{1})c_{212}(u_{1},u_{2})+b_{13}(u,u_{1})G_{d_{2}}(u_{1},u_{2})
\nonumber \\
&&+\omega (u_{1},u)a_{11}(u,u_{1})a_{13}(u,u_{2})G_{d_{2}}(u,u_{1})
\label{baba.57}
\end{eqnarray}%
and 
\begin{eqnarray}
H_{j1}(u_{1},u_{2}) &=&a_{j6}(u,u_{1})\left(
c_{16}(u_{1},u_{2})+c_{18}(u_{1},u_{2})\right) +a_{j7}(u,u_{1})\left(
c_{26}(u_{1},u_{2})+c_{29}(u_{1},u_{2})\right)  \nonumber \\
&&+b_{j2}(u,u_{1})G_{d_{1}}(u_{1},u_{2})+\omega
(u_{1},u)a_{j1}(u,u_{1})a_{j2}(u,u_{2})G_{d_{1}}(u,u_{1})  \nonumber \\
&&+a_{j1}(u,u_{1})a_{j4}(u,u_{2})d_{13}(u_{1},u)  \nonumber \\
H_{j2}(u_{1},u_{2}) &=&a_{j6}(u,u_{1})\left(
c_{17}(u_{1},u_{2})+c_{110}(u_{1},u_{2})\right) +a_{j7}(u,u_{1})\left(
c_{27}(u_{1},u_{2})+c_{211}(u_{1},u_{2})\right)  \nonumber \\
&&+b_{j3}(u,u_{1})G_{d_{1}}(u_{1},u_{2})+\omega
(u_{1},u)a_{j1}(u,u_{1})a_{j2}(u,u_{2})G_{d_{2}}(u,u_{1})  \nonumber \\
&&+a_{j1}(u,u_{1})a_{j4}(u,u_{2})d_{14}(u_{1},u)  \nonumber \\
H_{j3}(u_{1},u_{2})
&=&a_{j6}(u,u_{1})c_{19}(u_{1},u_{2})+a_{j7}(u,u_{1})c_{210}(u_{1},u_{2})+b_{j2}(u,u_{1})G_{d_{2}}(u_{1},u_{2})
\nonumber \\
&&+\omega
(u_{1},u)a_{j1}(u,u_{1})a_{j3}(u,u_{2})G_{d_{1}}(u,u_{1})+a_{j1}(u,u_{1})a_{j5}(u,u_{2})d_{13}(u_{1},u)
\nonumber \\
H_{j4}(u_{1},u_{2})
&=&a_{j6}(u,u_{1})c_{111}(u_{1},u_{2})+a_{j7}(u,u_{1})c_{212}(u_{1},u_{2})+b_{j3}(u,u_{1})G_{d_{2}}(u_{1},u_{2})
\nonumber \\
&&+\omega
(u_{1},u)a_{j1}(u,u_{1})a_{j3}(u,u_{2})G_{d_{2}}(u,u_{1})+a_{j1}(u,u_{1})a_{j5}(u,u_{2})d_{14}(u_{1},u)
\label{baba.58}
\end{eqnarray}%
for $j=2,3$.

Again, $\Psi _{2}(u_{1},u_{2})$ will be an eigenstate of $t(u)$ with
eigenvalue%
\begin{equation}
\Lambda _{2}(u,u_{1},u_{2})=\sum_{j=1}^{3}\Omega _{j}(u){\cal X}%
_{j}(u)a_{j1}(u,u_{1})a_{j1}(u,u_{2})  \label{baba.59}
\end{equation}%
provided the following equations are satisfied%
\begin{equation}
\frac{{\cal X}_{1}(u_{1})}{{\cal X}_{2}(u_{1})}=\Theta (u,u_{1})\frac{%
a_{21}(u_{1},u_{2})}{a_{11}(u_{1},u_{2})},\qquad \frac{{\cal X}_{1}(u_{2})}{%
{\cal X}_{2}(u_{2})}=\Theta (u,u_{2})\frac{a_{21}(u_{2},u_{1})}{%
a_{11}(u_{2},u_{1})}\qquad  \label{baba.60}
\end{equation}%
where $\Theta (u_{i}),\ i=1,2$ are given by\ (\ref{baba.41}).

\subsection{The n-particle state}

From the previous results we can are seek for operator valued functions with
a recurrence relation of the form%
\begin{eqnarray}
\Phi _{n}(u,\ldots ,u_{n}) &=&{\cal B}_{1}(u_{1})\Phi _{n-1}(u_{2},\ldots
,u_{n})  \nonumber \\
&&+{\cal B}_{2}(u_{1})\sum_{i=2}^{n}\digamma _{1}^{(i)}(u_{1},\ldots
,u_{n})\Phi _{n-2}(u_{2},\ldots ,\overset{\vee }{u_{i}},\ldots ,u_{n}){\cal D%
}_{1}(u_{i})  \nonumber \\
&&+{\cal B}_{2}(u_{1})\sum_{i=2}^{n}\digamma _{2}^{(i)}(u_{1},\ldots
,u_{n})\Phi _{n-2}(u_{2},\ldots ,\overset{\vee }{u_{i}},\ldots ,u_{n}){\cal D%
}_{2}(u_{i})  \label{baba.61}
\end{eqnarray}%
It was shown in \cite{GLL} that the above operator is normal ordered
satisfying $n-1$ exchange conditions%
\begin{equation}
\Phi _{n}(u_{1},\ldots ,u_{i},u_{i+1},\ldots ,u_{n})=\omega
(u_{i},u_{i+1})\Phi _{n}(u_{1},\ldots ,u_{i+1},u_{i},\ldots ,u_{n})
\label{baba.62}
\end{equation}%
provided the functions $\digamma _{\alpha }^{(i)}(u_{1},\ldots ,u_{n})$ are
given by 
\begin{equation}
\digamma _{\alpha }^{(i)}(u_{1},\ldots
,u_{n})=\dprod\limits_{j=2}^{i-1}\omega (u_{j},u_{i})\dprod_{k=2,k\neq
i}^{n}a_{\alpha 1}(u_{i},u_{k})G_{d_{\alpha }}(u_{1},u_{i}),\qquad (\alpha
=1,2)  \label{baba.63}
\end{equation}%
Therefore the $n$-particle state will be given by%
\begin{equation}
\Psi _{n}(u_{1},\ldots ,u_{n})=\Phi _{n}(u,\ldots ,u_{n})\left\vert
0\right\rangle  \label{baba.64}
\end{equation}%
and the action of the operators ${\cal D}_{\alpha }(u)$, $\alpha =1,2,3$, on
this state will be represented by 
\[
{\cal D}_{\alpha }(u)\Psi _{n}(u_{1},\ldots ,u_{n})={\cal X}_{\alpha
}(u)\dprod\limits_{i=1}^{n}a_{\alpha 1}(u,u_{i})\Psi _{n}(u_{1},\ldots
,u_{n}) 
\]%
\[
+\sum_{i=1}^{n}\dprod\limits_{j=1}^{i-1}\omega (u_{j},u_{i})[{\cal X}%
_{1}(u)a_{\alpha 2}(u,u_{i})\dprod\limits_{j\neq i}^{n}a_{11}(u,u_{j})+{\cal %
X}_{2}(u)a_{\alpha 3}(u,u_{i})\dprod\limits_{j\neq i}^{n}a_{21}(u,u_{j})]%
{\cal B}_{1}(u)\Psi _{n-1}(\overset{\vee }{u_{i}}) 
\]%
\[
+(1-\delta _{\alpha ,1})\sum_{i=1}^{n}\dprod\limits_{j=1}^{i-1}\omega
(u_{j},u_{i})[{\cal X}_{1}(u)a_{\alpha 4}(u,u_{i})\dprod\limits_{j\neq
i}^{n}a_{11}(u,u_{j})+{\cal X}_{2}(u)a_{\alpha
5}(u,u_{i})\dprod\limits_{j\neq i}^{n}a_{21}(u,u_{j})]{\cal B}_{3}(u)\Psi
_{n-1}(\overset{\vee }{u_{i}}) 
\]%
\begin{eqnarray}
&&+\sum_{i=1}^{n-1}\sum_{j=i+1}^{n}\left\{ {\cal X}_{1}(u_{i}){\cal X}%
_{1}(u_{j})\dprod_{k\neq i,j}^{n}a_{11}(u_{i},u_{k})\dprod_{l\neq
i,j}^{n}a_{11}(u_{j},u_{l})H_{\alpha 1}(u_{i},u_{j})\right.  \nonumber \\
&&+{\cal X}_{2}(u_{i}){\cal X}_{1}(u_{j})\dprod_{k\neq
i,j}^{n}a_{21}(u_{i},u_{k})\dprod_{l\neq
i,j}^{n}a_{11}(u_{j},u_{l})H_{\alpha 2}(u_{i},u_{j})  \nonumber \\
&&+{\cal X}_{1}(u_{i}){\cal X}_{2}(u_{j})\dprod_{k\neq
i,j}^{n}a_{11}(u_{i},u_{k})\dprod_{l\neq
i,j}^{n}a_{21}(u_{j},u_{l})H_{\alpha 3}(u_{i},u_{j})  \nonumber \\
&&\left. +{\cal X}_{2}(u_{i}){\cal X}_{2}(u_{j})\dprod_{k\neq
i,j}^{n}a_{21}(u_{i},u_{k})\dprod_{l\neq
i,j}^{n}a_{21}(u_{j},u_{l})H_{\alpha 4}(u_{i},u_{j})\right\}  \nonumber \\
&&\times \dprod\limits_{k=1}^{i-1}\omega (u_{k},u_{i})\dprod\limits_{l=1\neq
i}^{i-1}\omega (u_{l},u_{j}){\cal B}_{2}(u)\Psi _{n-2}(\overset{\vee }{u_{i}}%
,\overset{\vee }{u_{j}})  \label{baba.65}
\end{eqnarray}

Finally, the corresponding $n$-particle eigenvalue problem will be%
\begin{equation}
t(u)\Psi _{n}(u_{1},\ldots ,u_{n})=\left( \sum_{\alpha =1}^{3}\Omega
_{\alpha }(u){\cal X}_{\alpha }(u)\dprod\limits_{i=1}^{n}a_{\alpha
1}(u,u_{i})\right) \Psi _{n}(u_{1},\ldots ,u_{n})  \label{baba.66}
\end{equation}%
provided that the {\small BA} equations are satisfied%
\begin{equation}
\frac{{\cal X}_{1}(u_{k})}{{\cal X}_{2}(u_{k})}=\Theta
(u_{k})\dprod\limits_{j=1,j\neq k}^{n}\frac{a_{21}(u_{k},u_{j})}{%
a_{11}(u_{k},u_{j})},\qquad (k=1,2,\ldots ,n)  \label{baba.67}
\end{equation}

\section{Explicit solutions}

In this section explicit expressions of the eigenvalue problem are presented
for both models,. First we recall the fundamental relation (\ref{baba.35})
to get the coefficients $a_{ij}(u,v)$ which appear effectively in the 
{\small BA} expressions (\ref{baba.66}) and (\ref{baba.67}):%
\begin{eqnarray}
a_{11}(u,v) &=&\frac{x_{1}(u-v)}{x_{2}(u-v)}\frac{x_{2}(u+v)}{x_{1}(u+v)} 
\nonumber \\
a_{21}(u,v) &=&-\omega (u,v)[\frac{x_{1}(u+v)x_{4}(u+v)+x_{5}(u+v)y_{5}(u+v)%
}{x_{1}(u+v)x_{2}(u+v)}]  \nonumber \\
a_{31}(u,v) &=&\frac{x_{2}(u-v)}{x_{3}(u-v)}[\frac{%
x_{2}(u+v)^{2}+x_{6}(u+v)y_{6}(u+v)}{x_{2}(u+v)x_{3}(u+v)}]  \label{exsol.1}
\end{eqnarray}%
For the factor with the boundary contributions (\ref{baba.41}) we will
consider only the expression%
\begin{equation}
\Theta (u_{i})=-\frac{\Omega _{2}(u)a_{25}(u,u_{i})+\Omega
_{3}(u)a_{35}(u,u_{i})}{\Omega _{2}(u)a_{24}(u,u_{i})+\Omega
_{3}(u)a_{34}(u,u_{i})}  \label{exsol.2}
\end{equation}%
where the $\Omega _{j}(u)$ are given by\ (\ref{baba.21}) and%
\begin{eqnarray}
a_{24}(u,v) &=&-[\frac{x_{6}(u-v)}{x3(u-v)}\frac{x_{3}(u+v)}{x_{2}(u+v)}%
-f_{1}(v)\frac{x_{6}(u+v)}{x_{2}(u+v)}]  \nonumber \\
a_{25}(u,v) &=&\frac{x_{6}(u+v)}{x_{2}(u+v)}  \nonumber \\
a_{34}(u,v) &=&-f_{3}(u)[f_{1}(v)\frac{x_{6}(u+v)}{x_{2}(u+v)}-\frac{%
x_{6}(u-v)}{x_{3}(u-v)}\frac{x_{3}(u+v)}{x_{2}(u+v)}]  \nonumber \\
&&+f_{1}(v)\frac{y_{6}(u-v)}{x_{3}(u-v)}[\frac{%
x_{6}(u+v)y_{6}(u+v)+x_{2}^{2}(u+v)}{x_{2}(u+v)x_{3}(u+v)}]  \nonumber \\
&&-[\frac{x_{6}(u-v)y_{6}(u-v)+x_{2}^{2}(u-v)}{x_{3}^{2}(u-v)}]\frac{%
y_{6}(u+v)}{x_{2}(u+v)}  \nonumber \\
a_{35}(u,v) &=&-f_{3}(u)\frac{x_{6}(u+v)}{x_{2}(u+v)}+\frac{y_{6}(u-v)}{%
x_{3}(u-v)}[\frac{x_{6}(u+v)y_{6}(u+v)+x_{2}^{2}(u+v)}{x_{2}(u+v)x_{3}(u+v)}]
\label{exsol.3}
\end{eqnarray}%
with the $f_{i}(u)$ given by (\ref{baba.13}).

\subsection{sl(2\TEXTsymbol{\vert}1)$^{(2)}$ model}

Substituting the matrix elements of the ${\cal R}$ matrix (\ref{mod.5}) and
the matrix elements of the $K$ matrices (\ref{mod.7}) and (\ref{mod.9}) for
this model we get the following expressions in our algebraic {\small BA}: \
first the coefficients $a_{j1}(u,u_{i})$ for the eigenvalue $\Lambda
_{n}(u,\{u_{i}\})$

\begin{equation}
a_{11}(u,u_{i})=\frac{\sinh (u+u_{i})}{\sinh (u+u_{i}+2\eta )}\frac{\sinh
(u-u_{i}-2\eta )}{\sinh (u-u_{i})},  \label{exsol.4}
\end{equation}%
\begin{equation}
a_{21}(u,u_{i})=\frac{\sinh (u+u_{i})}{\sinh (u+u_{i}+2\eta )}\frac{\sinh
(u-u_{i}-2\eta )}{\sinh (u-u_{i})}\frac{\cosh (u+u_{i}+3\eta )}{\cosh
(u+u_{i}+\eta )}\frac{\cosh (u-u_{i}+\eta )}{\cosh (u-u_{i}-\eta )},
\label{exsol.5}
\end{equation}%
\[
a_{31}(u,u_{i})=\frac{\cosh (u+u_{i}+3\eta )}{\cosh (u+u_{i}+\eta )}\frac{%
\cosh (u-u_{i}+\eta )}{\cosh (u-u_{i}-\eta )}, 
\]%
Second, the factors of $\Lambda _{n}(u,\{u_{i}\})$ and of the {\small BA}
equations with boundary contributions%
\begin{equation}
\Omega _{1}(u)=\frac{\cosh (2u+3\eta )}{\cosh (2u+\eta )}\frac{\alpha \sinh
u-2\cosh u}{\alpha \sinh (u+2\eta )-2\cosh (u+2\eta )}\frac{\alpha \cosh
(u+\eta )-2\sinh (u+\eta )}{\alpha \cosh (u+\eta )+2\sinh (u+\eta )},
\label{exsol.6}
\end{equation}%
\begin{equation}
\Omega _{2}(u)=-\frac{\sinh (2u+2\eta )}{\sinh (2u)}\frac{\alpha \sinh
u-2\cosh u}{\alpha \sinh (u+2\eta )-2\cosh (u+2\eta )},  \label{exsol.7}
\end{equation}%
\begin{equation}
\Omega _{3}(u)=-\frac{\alpha \sinh u-2\cosh u}{\alpha \sinh (u+2\eta
)-2\cosh (u+2\eta )},  \label{exsol.8}
\end{equation}%
\begin{equation}
{\cal X}_{1}(u)=-\frac{\beta \sinh u+2\cosh u}{\beta \sinh u-2\cosh u}\frac{%
x_{1}^{2N}(u)}{f^{N}(u)},  \label{exsol.9}
\end{equation}%
\begin{equation}
{\cal X}_{2}(u)=\frac{\sinh (2u)}{\sinh (2u+2\eta )}\frac{\beta \sinh
(u+2\eta )-2\cosh (u+2\eta )}{\beta \sinh u-2\cosh u}\frac{x_{2}^{2N}(u)}{%
f^{N}(u)},  \label{exsol.10}
\end{equation}%
\begin{equation}
{\cal X}_{3}(u)=\frac{\cosh (2u-\eta )}{\cosh (2u+\eta )}\frac{\beta \sinh
(u+2\eta )-2\cosh (u+2\eta )}{\beta \sinh u-2\cosh u}\frac{\beta \cosh
(u+\eta )-2\sinh (u+\eta )}{\beta \cosh (u-\eta )+2\sinh (u-\eta )}\frac{%
x_{3}^{2N}(u)}{f^{N}(u)},  \label{exsol.11}
\end{equation}%
and 
\begin{equation}
\Theta (u_{i})=\frac{\sinh (2u_{i}+2\eta )}{\sinh (2u_{i})}\frac{\alpha
\cosh (u_{i}+\eta )+2\sinh (u_{i}+\eta )}{\alpha \cosh (u_{i}+\eta )-2\sinh
(u_{i}+\eta )}.  \label{exsol.12}
\end{equation}

From these data one can see that the n-particle state $\Psi _{n}(\{u_{i}\})$
is an eigenfuction of the transfer matrix (\ref{baba.20}) \ for the $%
sl(2|1)^{(2)}$ vertex model with eigenvalue%
\begin{eqnarray}
\Lambda _{n}(u,\{u_{i}\}) &=&\Omega _{1}(u){\cal X}_{1}(u)\dprod%
\limits_{i=1}^{n}\frac{\sinh (u+u_{i}-\eta )}{\sinh (u+u_{i}+\eta )}\frac{%
\sinh (u-u_{i}-\eta )}{\sinh (u-u_{i}+\eta )}  \nonumber \\
&&-\Omega _{2}(u){\cal X}_{2}(u)\dprod\limits_{i=1}^{n}\frac{\sinh
(u+u_{i}-\eta )}{\sinh (u+u_{i}+\eta )}\frac{\sinh (u-u_{i}-\eta )}{\sinh
(u-u_{i}+\eta )}  \nonumber \\
&&\times \frac{\cosh (u+u_{i}+2\eta )}{\cosh (u+u_{i})}\frac{\cosh
(u-u_{i}+2\eta )}{\cosh (u-u_{i})}  \nonumber \\
&&+\Omega _{3}(u){\cal X}_{3}(u)\dprod\limits_{i=1}^{n}\frac{\cosh
(u+u_{i}+2\eta )}{\cosh (u+u_{i})}\frac{\cosh (u-u_{i}+2\eta )}{\cosh
(u-u_{i})}  \label{exsol.13}
\end{eqnarray}%
provided that the parameters $\{u_{i}\}$ satisfy the {\small BA} equations%
\begin{eqnarray}
\left( \frac{\sinh (u_{i}+\eta }{\sinh (u_{i}-\eta )}\right) ^{2N} &=&-\frac{%
\alpha \cosh u_{i}+2\sinh u_{i}}{\alpha \cosh u_{i}-2\sinh u_{i}}\frac{\beta
\sinh (u_{i}+\eta )-2\cosh (u_{i}+\eta )}{\beta \sinh (u_{i}-\eta )-2\cosh
(u_{i}-\eta )}  \nonumber \\
&&\times \dprod\limits_{\{j\neq i\}=1}^{n}\frac{\cosh (u_{i}+u_{j}+\eta )}{%
\cosh (u_{i}+u_{j}-\eta )}\frac{\cosh (u_{i}-u_{j}+\eta )}{\cosh
(u_{i}-u_{j}-\eta )}  \nonumber \\
i &=&1,2,...,n  \label{exsol.15}
\end{eqnarray}%
where we have used the shifts $u_{i}\rightarrow u_{i}=u_{i}-\eta $ to bring
these expressions into a symmetric form in $\eta .$

The formulation of this model in terms of the {\small QISM} presented here
is new. However, one can verify that our results give the energy
eigenspectrum previously obtained in the framework of coordinate {\small BA}
by Fireman {\it et al.} \cite{FLU}.

\subsection{osp(2\TEXTsymbol{\vert}1) model}

For this model the $K$ matrices have no free parameters but we have to
consider three cases. For both cases the coefficients $a_{j1}(u,u_{i})$ are
given by%
\begin{eqnarray}
a_{11}(u,u_{i}) &=&\frac{\sin (u+u_{i})}{\sinh (u+u_{i}+2\eta )}\frac{\sinh
(u-u_{i}-2\eta )}{\sinh (u-u_{i})},  \nonumber \\
a_{21}(u,u_{i}) &=&\frac{\sin (u+u_{i}+4\eta )}{\sinh (u+u_{i}+3\eta )}\frac{%
\sinh (u+u_{i}+\eta )}{\sinh (u+u_{i}+2\eta )}\frac{\sin (u-u_{i}+2\eta )}{%
\sinh (u-u_{i})}\frac{\sinh (u-u_{i}-\eta )}{\sinh (u-u_{i}+\eta )}, 
\nonumber \\
a_{31}(u,u_{i}) &=&\frac{\sinh (u+u_{i}+5\eta )}{\sinh (u+u_{i}+3\eta )}%
\frac{\sinh (u-u_{i}+3\eta )}{\sinh (u-u_{i}+\eta )}.  \label{exsol.16}
\end{eqnarray}%
Though, the boundary contributions are different for each cases:

\subsubsection{The (1,M) case}

In this case we have%
\begin{eqnarray}
\Omega _{1}(u) &=&\frac{\sinh (2u+\eta )}{\sinh (2u+2\eta )}\frac{\sinh
(2u+6\eta )}{\sinh (2u+3\eta )}  \nonumber \\
\Omega _{2}(u) &=&-{\rm e}^{2\eta }\frac{\sinh (2u+6\eta )}{\sinh (2u+4\eta )%
}  \nonumber \\
\Omega _{3}(u) &=&{\rm e}^{2\eta }  \label{exsol.17}
\end{eqnarray}%
\begin{eqnarray}
{\cal X}_{1}(u) &=&\frac{x_{1}^{2N}(u)}{f^{N}(u)}  \nonumber \\
{\cal X}_{2}(u) &=&{\rm e}^{-2\eta }\frac{\sinh (2u)}{\sinh (2u+2\eta )}%
\frac{x_{2}^{2N}(u)}{f^{N}(u)}  \nonumber \\
{\cal X}_{3}(u) &=&{\rm e}^{-2\eta }\frac{\sinh (2u)}{\sinh (2u+4\eta )}%
\frac{\sinh (2u+5\eta )}{\sinh (2u+3\eta )}\frac{x_{3}^{2N}(u)}{f^{N}(u)}
\label{exsol.18}
\end{eqnarray}%
and%
\begin{equation}
\Theta (u_{i})={\rm e}^{2\eta }\frac{\sinh (2u_{i}+2\eta )}{\sinh (2u_{i})}%
,\qquad i=1,...,n  \label{exsol.19}
\end{equation}

Therefore, the $n$-particle state $\Psi _{n}(\{u_{i}\})$ is an eigenfuction
of the transfer matrix (\ref{baba.20}) \ for the $osp(2|1)$ vertex model
with boundaries $(1,M)$. The corresponding eigenvalue is given by%
\begin{eqnarray}
\Lambda _{n}(u,\{u_{i}\}) &=&\frac{\sinh (2u+\eta )}{\sinh (2u+2\eta )}\frac{%
\sinh (2u+6\eta )}{\sinh (2u+3\eta )}\frac{x_{1}^{2N}(u)}{f^{N}(u)}%
\dprod\limits_{i=1}^{n}\frac{\sin (u+u_{i})}{\sinh (u+u_{i}+2\eta )}\frac{%
\sinh (u-u_{i}-2\eta )}{\sinh (u-u_{i})}  \nonumber \\
&&-\frac{\sinh (2u+6\eta )}{\sinh (2u+4\eta )}\frac{\sinh (2u)}{\sinh
(2u+2\eta )}\frac{x_{2}^{2N}(u)}{f^{N}(u)}\dprod\limits_{i=1}^{n}\frac{\sin
(u+u_{i}+4\eta )}{\sinh (u+u_{i}+3\eta )}\frac{\sinh (u+u_{i}+\eta )}{\sinh
(u+u_{i}+2\eta )}  \nonumber \\
&&\times \frac{\sin (u-u_{i}+2\eta )}{\sinh (u-u_{i})}\frac{\sinh
(u-u_{i}-\eta )}{\sinh (u-u_{i}+\eta )}  \nonumber \\
&&+\frac{\sinh (2u)}{\sinh (2u+4\eta )}\frac{\sinh (2u+5\eta )}{\sinh
(2u+3\eta )}\frac{x_{3}^{2N}(u)}{f^{N}(u)}\dprod\limits_{i=1}^{n}\frac{\sinh
(u+u_{i}+5\eta )}{\sinh (u+u_{i}+3\eta )}\frac{\sinh (u-u_{i}+3\eta )}{\sinh
(u-u_{i}+\eta )}  \nonumber \\
&&  \label{exsol.20}
\end{eqnarray}%
provided that its parameters $\{u_{i}\}$are solutions of the {\small BA}
equations%
\begin{eqnarray}
\left( \frac{\sinh (u_{i}+2\eta )}{\sinh u_{i}}\right) ^{2N}
&=&\dprod\limits_{\{j\neq i\}=1}^{n}\frac{\sin (u_{i}+u_{j}+4\eta )}{\sinh
(u_{i}+u_{j}+3\eta )}\frac{\sinh (u_{i}+u_{j}+\eta )}{\sin (u_{i}+u_{j})}%
\frac{\sin (u_{i}-u_{j}+2\eta )}{\sinh (u_{i}-u_{j}-2\eta )}\frac{\sinh
(u_{i}-u_{j}-\eta )}{\sinh (u_{i}-u_{j}+\eta )}  \nonumber \\
i &=&1,..,n  \label{exsol.21}
\end{eqnarray}

\subsubsection{The (F$^{+}$,G$^{+}$) case}

Here we have%
\begin{eqnarray}
\Omega _{1}(u) &=&-{\rm e}^{2u}\frac{\sinh (2u+6\eta )}{\sinh (2u+2\eta )}%
\frac{\sinh (u+\frac{5}{2}\eta )}{\sinh (u+\frac{9}{2}\eta )}\frac{\cosh (u+%
\frac{1}{2}\eta )}{\cosh (u+\frac{3}{2}\eta )}  \nonumber \\
\Omega _{2}(u) &=&-\frac{\sinh (2u+6\eta )}{\sinh (2u+4\eta )}\frac{\sinh (u+%
\frac{5}{2}\eta )}{\sinh (u+\frac{9}{2}\eta )}  \nonumber \\
\Omega _{3}(u) &=&-{\rm e}^{-2u-4\eta }\frac{\sinh (u+\frac{3}{2}\eta )}{%
\sinh (u+\frac{9}{2}\eta )}  \label{exsol.22}
\end{eqnarray}%
\begin{eqnarray}
{\cal X}_{1}(u) &=&-{\rm e}^{-2u}\frac{\sinh (u+\frac{3}{2}\eta )}{\sinh (u-%
\frac{3}{2}\eta )}\frac{x_{1}^{2N}(u)}{f^{N}(u)}  \nonumber \\
{\cal X}_{2}(u) &=&\frac{\sinh (2u)}{\sinh (2u+2\eta )}\frac{\sinh (u+\frac{1%
}{2}\eta )}{\sinh (u-\frac{3}{2}\eta )}\frac{x_{2}^{2N}(u)}{f^{N}(u)} 
\nonumber \\
{\cal X}_{3}(u) &=&-{\rm e}^{2u+4\eta }\frac{\sinh (2u)}{\sinh (2u+4\eta )}%
\frac{\sinh (u+\frac{1}{2}\eta )}{\sinh (u-\frac{3}{2}\eta )}\frac{\cosh (u+%
\frac{5}{2}\eta )}{\cosh (u+\frac{3}{2}\eta )}\frac{x_{3}^{2N}(u)}{f^{N}(u)}
\label{exsol.23}
\end{eqnarray}%
and%
\begin{equation}
\Theta (u_{i})=-{\rm e}^{-2u_{i}}\frac{\sinh (2u_{i}+2\eta )}{\sinh (2u_{i})}%
\frac{\sinh (u_{i}+\frac{1}{2}\eta )}{\sinh (u_{i}+\frac{3}{2}\eta )}
\label{exsol.24}
\end{equation}%
Therefore, the $n$-particle state $\Psi _{n}(\{u_{i}\})$ is an eigenfuction
of the transfer matrix (\ref{baba.20}) \ for the $osp(2|1)$ vertex model
with boundaries $(F^{+},G^{+})$. The corresponding eigenvalue is given by%
\begin{eqnarray}
\Lambda _{n}(u,\{u_{i}\}) &=&\Omega _{1}(u){\cal X}_{1}(u)\dprod%
\limits_{i=1}^{n}\frac{\sin (u+u_{i}-\eta )}{\sinh (u+u_{i}+\eta )}\frac{%
\sinh (u-u_{i}-\eta )}{\sinh (u-u_{i}+\eta )}  \nonumber \\
&&+\Omega _{2}(u){\cal X}_{2}(u)\dprod\limits_{i=1}^{n}\frac{\sin
(u+u_{i}+3\eta )}{\sinh (u+u_{i}+2\eta )}\frac{\sin (u-u_{i}+3\eta )}{\sinh
(u-u_{i}+\eta )}  \nonumber \\
&&\times \frac{\sinh (u+u_{i})}{\sinh (u+u_{i}+\eta )}\frac{\sinh (u-u_{i})}{%
\sinh (u-u_{i}+2\eta )}  \nonumber \\
&&+\Omega _{3}(u){\cal X}_{3}(u)\dprod\limits_{i=1}^{n}\frac{\sinh
(u+u_{i}+4\eta )}{\sinh (u+u_{i}+2\eta )}\frac{\sinh (u-u_{i}+4\eta )}{\sinh
(u-u_{i}+2\eta )}  \label{exsol.25}
\end{eqnarray}%
with the {\small BA} equations%
\begin{eqnarray}
\left( \frac{\sinh (u_{i}+\eta )}{\sinh (u_{i}-\eta )}\right) ^{2N}
&=&\left( \frac{\sinh (u_{i}-\frac{1}{2}\eta )}{\sinh (u_{i}+\frac{1}{2}\eta
)}\right) ^{2}\dprod\limits_{\{j\neq i\}=1}^{n}\frac{\sin (u_{i}+u_{j}+2\eta
)}{\sinh (u_{i}+u_{j}+\eta )}\frac{\sin (u_{i}-u_{j}+2\eta )}{\sinh
(u_{i}-u_{j}-2\eta )}  \nonumber \\
&&\times \frac{\sinh (u_{i}+u_{j}-\eta )}{\sin (u_{i}+u_{j}+\eta )}\frac{%
\sinh (u_{i}-u_{j}-\eta )}{\sinh (u_{i}-u_{j}+\eta )}  \nonumber \\
i &=&1,...,n  \label{exsol.26}
\end{eqnarray}%
Again, $\Lambda _{n}(u,\{u_{i}\})$ and the Bethe equations have been written
in their symmetric form.

\subsubsection{The (F$^{-}$,G$^{-}$) case}

Here we have%
\begin{eqnarray}
\Omega _{1}(u) &=&{\rm e}^{2u}\frac{\sinh (2u+6\eta )}{\sinh (2u+2\eta )}%
\frac{\cosh (u+\frac{5}{2}\eta )}{\cosh (u+\frac{9}{2}\eta )}\frac{\sinh (u+%
\frac{1}{2}\eta )}{\sinh (u+\frac{3}{2}\eta )}  \nonumber \\
\Omega _{2}(u) &=&-\frac{\sinh (2u+6\eta )}{\sinh (2u+4\eta )}\frac{\cosh (u+%
\frac{5}{2}\eta )}{\cosh (u+\frac{9}{2}\eta )}  \nonumber \\
\Omega _{3}(u) &=&{\rm e}^{-2u-4\eta }\frac{\cosh (u+\frac{3}{2}\eta )}{%
\cosh (u+\frac{9}{2}\eta )}  \label{exsol.27}
\end{eqnarray}%
\begin{eqnarray}
{\cal X}_{1}(u) &=&{\rm e}^{-2u}\frac{\cosh (u+\frac{3}{2}\eta )}{\cosh (u-%
\frac{3}{2}\eta )}\frac{x_{1}^{2N}(u)}{f^{N}(u)}  \nonumber \\
{\cal X}_{2}(u) &=&\frac{\sinh (2u)}{\sinh (2u+2\eta )}\frac{\cosh (u+\frac{1%
}{2}\eta )}{\cosh (u-\frac{3}{2}\eta )}\frac{x_{2}^{2N}(u)}{f^{N}(u)} 
\nonumber \\
{\cal X}_{3}(u) &=&{\rm e}^{2u+4\eta }\frac{\sinh (2u)}{\sinh (2u+4\eta )}%
\frac{\cosh (u+\frac{1}{2}\eta )}{\cosh (u-\frac{3}{2}\eta )}\frac{\sinh (u+%
\frac{5}{2}\eta )}{\sinh (u+\frac{3}{2}\eta )}\frac{x_{3}^{2N}(u)}{f^{N}(u)}
\label{exsol.28}
\end{eqnarray}%
\begin{equation}
\Theta (u_{i})={\rm e}^{-2u_{i}}\frac{\sinh (2u_{i}+2\eta )}{\sinh (2u_{i})}%
\frac{\cosh (u_{i}+\frac{1}{2}\eta )}{\cosh (u_{i}+\frac{3}{2}\eta )}
\label{exsol.29}
\end{equation}%
Therefore, the $n$-particle state $\Psi _{n}(\{u_{i}\})$ is an eigenfuction
of the transfer matrix (\ref{baba.20}) \ for the $osp(2|1)$ vertex model
with boundaries $(F^{-},G^{-})$. The corresponding eigenvalue is given by%
\begin{eqnarray}
\Lambda _{n}(u,\{u_{i}\}) &=&\Omega _{1}(u){\cal X}_{1}(u)\dprod%
\limits_{i=1}^{n}\frac{\sin (u+u_{i}-\eta )}{\sinh (u+u_{i}+\eta )}\frac{%
\sinh (u-u_{i}-\eta )}{\sinh (u-u_{i}+\eta )}  \nonumber \\
&&+\Omega _{2}(u){\cal X}_{2}(u)\dprod\limits_{i=1}^{n}\frac{\sin
(u+u_{i}+3\eta )}{\sinh (u+u_{i}+2\eta )}\frac{\sin (u-u_{i}+3\eta )}{\sinh
(u-u_{i}+\eta )}  \nonumber \\
&&\times \frac{\sinh (u+u_{i})}{\sinh (u+u_{i}+\eta )}\frac{\sinh (u-u_{i})}{%
\sinh (u-u_{i}+2\eta )}  \nonumber \\
&&+\Omega _{3}(u){\cal X}_{3}(u)\dprod\limits_{i=1}^{n}\frac{\sinh
(u+u_{i}+4\eta )}{\sinh (u+u_{i}+2\eta )}\frac{\sinh (u-u_{i}+4\eta )}{\sinh
(u-u_{i}+2\eta )}  \label{exsol.30}
\end{eqnarray}%
and the {\small BA} equations are now given by%
\begin{eqnarray}
\left( \frac{\sinh (u_{i}+\eta )}{\sinh (u_{i}-\eta )}\right) ^{2N}
&=&\left( \frac{\cosh (u_{i}-\frac{1}{2}\eta )}{\cosh (u_{i}+\frac{1}{2}\eta
)}\right) ^{2}\dprod\limits_{\{j\neq i\}=1}^{n}\frac{\sin (u_{i}+u_{j}+2\eta
)}{\sinh (u_{i}+u_{j}+\eta )}\frac{\sin (u_{i}-u_{j}+2\eta )}{\sinh
(u_{i}-u_{j}-2\eta )}  \nonumber \\
&&\times \frac{\sinh (u_{i}+u_{j}-\eta )}{\sin (u_{i}+u_{j}-2\eta )}\frac{%
\sinh (u_{i}-u_{j}-\eta )}{\sinh (u_{i}-u_{j}+\eta )}  \label{exsol.31}
\end{eqnarray}%
Here, both $\Lambda _{n}(u,\{u_{i}\})$ and the {\small BA} equation were
written with $u_{i}\rightarrow u_{i}-\eta .$

These three cases were also considered in \cite{FLU} via the coordinate 
{\small BA}.

\section{Conclusion}

Here, with the aid of previous works \cite{GLL, KLS}, two of the three-state
graded $19$-vertex models have their boundary algebraic {\small BA} derived
using a generalization of the Tarasov's approach \cite{TA}. \ From our
results for the transfer matrix one can in principle derive the free-energy
thermodynamics, the quasi-particle excitations behavior as well as the
classes of universality governing the criticality of gapless regimes with
integrable boundary conditions. Moreover, the rather universal formula we
obtained for the eigenvectors could be useful in future computations of
off-shell properties such as form factors and correlation functions with
boundary conditions of relevant operators.

The algebraic {\small BA} for $n$-state models with periodic boundary
conditions was developed by Martins in \cite{MAR}. In a recent paper Galleas
and Martins \cite{GM} have presented the algebraic {\small BA} for the
vertex models based on superalgebras. Therefore we believe that the
Martins's approach can be generalized to include the diagonal open boundary
conditions.

Finally we observe that the vertex models discussed in this paper share a
common algebraic structure with the non-graded $19$-vertex models, the
Zamolodchikov-Fateev \cite{ZF} and the Izergin-Korepin \cite{IK} models.
Thus, we expected that a fusion procedure for the $sl(2|1)^{(2)}$ model \ as
well as the analytical {\small BA} formulation based in the quantum group
invariance of the $osp(2|1)$ model can reproduce our results.

{\bf Acknowledgment:} This work was supported in part by Funda\c{c}\~{a}o de
Amparo \`{a} Pesquisa do Estado de S\~{a}o Paulo-{\small FAPESP}-Brasil and
by Conselho Nacional de Desenvolvimento-{\small CNPq}-Brasil.

\end{document}